# Fluorescence photon measurements from single quantum dots on an optical nanofiber

**Ramachandrarao Yalla, K. P. Nayak*, and K. Hakuta**

*Center for Photonic Innovations, University of Electro-Communications, Chofu, Tokyo 182-8585, Japan*
*[kali@cpi.uec.ac.jp](mailto:kali@cpi.uec.ac.jp)*

**Abstract:** We experimentally investigate the fluorescence photon emission characteristics for single q-dots by using optical nanofibers. We demonstrate that single q-dots can be deposited along an optical nanofiber systematically and reproducibly with a precision of 5 μm. For single q-dots on an optical nanofiber, we measure the fluorescence photon numbers coupled into the nanofiber and the normalized photon correlations, by varying the excitation laser intensity. We estimate the fluorescence photon coupling efficiency into the nanofiber guided modes to be higher than $9.4\pm3\%$.

## 1. Introduction

Single photon manipulation is one of the major issues in the contemporary quantum optics, especially in the context of quantum information technology. For this purpose many novel ideas have been proposed so far. The key point of the ideas is to achieve strong confinement of the field using micro/nano boundary conditions. One major trend is to use various designs of high-$Q$ micro-structured resonators [1]. Recently, other than the micro-resonators, new type of nano-structured systems have been proposed and demonstrated. Examples would include plasmonic metal nano-wires [2] and sub-wavelength diameter silica fibers [3].

Regarding the sub-wavelength diameter silica fibers, termed as optical nanofibers, the research works are rapidly growing in various aspects in the last several years. It has been demonstrated using laser-cooled atoms that the atomic fluorescence can be channeled into the guided mode [4-6], and that single atoms can be readily detected by fluorescence measurement through nanofibers [3]. Photon correlations for atoms on a nanofiber have been systematically measured and analyzed by varying atom numbers from less than one to several [7]. Also it has been demonstrated that such a technique can be used for measuring fluorescence emission spectra from few atoms [8]. It has also been demonstrated that atoms can be trapped along the nanofiber using dipole-trapping method via propagating laser fields [9,10]. Furthermore, it has been theoretically predicted that the channeling efficiency of spontaneous emission into the nanofiber guided modes can be enhanced to 90% or higher by incorporating a cavity structure to the nanofiber, even with moderate finesse [11]. Recently, various nano-fabrication technologies have been successfully applied to micro/nano-size optical-fibers to create fiber Bragg gratings on them, and the cavity structures have been realized on nanofibers using the Bragg gratings [12,13].

Thus, the optical nanofiber method may open a promising way for manipulating atoms and photons. However, in order to extend the nanofiber method to real applications, such as single-photon source, one crucial issue would be to extend the photon emitters from atoms to solid-state emitters, such as semiconductor quantum dots (q-dots). In this paper, we experimentally investigate the fluorescence photon emission characteristics for single q-dots on an optical nanofiber using photon correlation spectroscopy. We use colloidal CdSeTe-nanocrystals as q-dots. We estimate the fluorescence photon coupling efficiency into the nanofiber guided modes to be higher than 9.4±3%. Regarding the technical aspect, we describe a method to deposit single q-dots systematically and reproducibly on optical nanofibers.

## 2. Experimental setup

Figure 1 illustrates the schematic diagram of the experimental setup. Main part of the setup consists of inverted microscope (Nikon, Eclipse Ti-U), optical nanofiber, and sub-pico-liter needle-dispenser (Applied Micro Systems, ND-2000). A high precision computer-controlled x-y stage is installed on the top of the microscope. The needle dispenser is fixed to another x-y stage on the microscope. The nanofiber is located at the central part of a tapered optical fiber produced by adiabatically tapering commercial single mode optical fibers using a heat and pull technique [6]. The diameter of the nanofiber is around 400 nm and is uniform for 2 mm along the fiber axis. The transmission through the tapered fiber is around 90%. The tapered fiber is installed into a cell with optical windows to protect from dusts. The cell is fixed on the computer-controlled x-y stage, and the central part of the tapered fiber (nanofiber region) is positioned at the focus of the microscope.

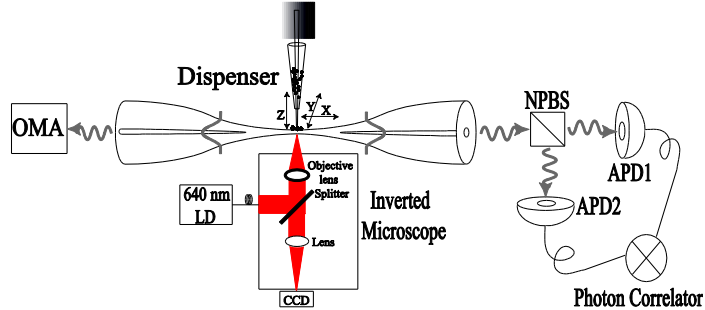

Fig. 1. (Color online) Schematic diagram of the experiment. The nanofiber is located at the central part of a tapered optical fiber. A sub-pico-liter needle-dispenser and an inverted microscope are used for depositing the q-dots on the nanofiber. The q-dots are excited using cw diode-laser at a wavelength of 640 nm. The fluorescence photons from q-dots coupled to the guided mode of the nanofiber are detected through the single mode optical fiber. At one end of the fiber the photon arrival times are recorded by using two-channel single-photon-counter, and at the other end the fluorescence emission spectrum is measured using optical multichannel analyzer (OMA). APD and NPBS denote avalanche-photodiode and non-polarizing beam splitter, respectively.

We use core-shell type colloidal CdSeTe(ZnS) q-dots having emission wavelength at 790 nm (Invitrogen, Q21371MP). The q-dot solution is diluted by pure water to reduce the concentration to $2\times10^{13}$ dots/cm$^3$. For depositing q-dots on the nanofiber, we use the sub-pico-liter needle-dispenser. The dispenser consists of a taper glass-tube which contains diluted q-dot solution and a needle having a tip of diameter 17 μm. The needle axis is adjusted to coincide with the axis of the microscope, and the needle tip position is computer-controlled along the z-axis. Once the needle tip passes through the taper glass-tube, it carries a small amount of q-dot solution at its edge. In order to deposit single/few q-dots on the nanofiber, the needle-tip position is adjusted so that the q-dot solution at its tip just touches the nanofiber. This is confirmed by sending a laser light through the nanofiber and observing the scattered light through the microscope.

Using such a technique q-dots are deposited periodically at 8 positions on the nanofiber by shifting the nanofiber along the x-axis in 20 μm steps. After depositing the q-dots, the residual solvent on the nanofiber is evaporated by flowing dust-free dry-nitrogen gas. The dry-nitrogen flow is maintained for whole experimental period of several days to protect both nanofiber and q-dots from dusts or oxygen gas. The transmission of the nanofiber is reduced to 81% after these procedures.

The q-dots are excited using cw diode-laser at a wavelength of 640 nm. The excitation beam is focused on to the nanofiber through the microscope objective lens with a spot size of 5 μm FWHM. The fluorescence photons emitted from q-dots are coupled to the guided mode of the nanofiber and are detected through the single mode optical fiber. At both ends of the fiber, the fluorescence light is filtered from the scattered excitation laser light by using a color glass filter (HOYA, R-72). At one end of the fiber, the fluorescence light beam is split into two using a 50:50 non-polarizing beam splitter (NPBS), and the split beams are re-coupled into multi-mode fibers and detected by fiber-coupled avalanche-photodiodes APD1 and APD2 (Perkin Elmer, SPCM-AQR/FC). Arrival times of photons at both APDs are recorded using a two-channel single-photon-counter (Pico Quant GmbH, Pico-Harp 300), and the arrival timing resolution is 300 ps including the response times of the APDs. The photon correlations are derived for the delay time up to 1 ms from the recorded arrival times for the two channels using offline program. Photon counts for each APD are obtained also from the records. At the other end of the fiber, the fluorescence emission spectrum is measured using an optical-multichannel-analyzer (Andor, OMA).

In order to estimate the absolute number of photons coupled into the guided modes, the light-transmission efficiency from the nanofiber region to the APD-detector would be important. We denote the efficiency as $\kappa$, and the value was measured to be $\kappa = 31\%$. The measured value is consistent with a value calculated as a product of transmission factors of optical nanofiber (81%), NPBS (51%), and color-glass filter (83%), and the coupling efficiency into the multi-mode fiber (90%). Note that we use the photon counts for APD1.

## 3. Results

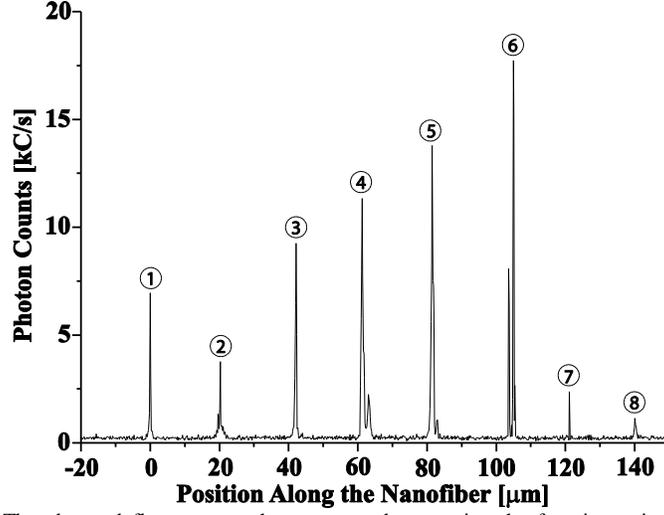

Fig. 2. The observed fluorescence photon counts by scanning the focusing point along the nanofiber. The excitation laser intensity was kept at 15 W/cm$^2$. One can clearly see the eight sharp signals along the nanofiber with a spacing of 20±5 μm, which well corresponds to the q-dot placement on the nanofiber. The signals are numbered from 1 to 8.

Figure 2 shows the observed fluorescence photon counts by scanning the focusing point along the nanofiber with a scanning speed of 2 μm/s. The excitation laser intensity was kept at 15 W/cm$^2$. One can clearly see the eight sharp signals along the nanofiber, with spacing of 20±5 μm, which well correspond to the q-dot placement on the nanofiber. The signals are numbered from 1 to 8. It is readily seen that each observed signal shows much sharper width than the focused beam size of 5 μm and some of the signals show double or triple peak structure. These signal behaviors may be understood due to blinking of q-dots [14,15] when the laser beam is passing over the q-dot position.

Fluorescence photon measurements are carried out for each signal position by varying the excitation laser intensity from 20 to 900 W/cm$^2$. Measurement time for each position is 5 min. Typical results for the photon counting and photon correlations are displayed in Fig. 3 for four positions (3, 4, 5, and 7). Excitation intensity is 50 W/cm$^2$ for the positions of 3, 4, and 5, and 130 W/cm$^2$ for the position 7. The left column shows the photon counts as a function of time. We average photon number over the time bin size of 17 ms and plot the histogram of fluorescence photons as a function of time. The right column shows the normalized photon correlations $g_N^{(2)}(\tau)$ for delay time ($\tau$) up to ±600 ns with a time resolution of 1 ns. The normalized correlations have been calculated by deriving the self-coincidences in each channel independently up to 1 ms.

Regarding the photon counts in the left column, the on/off-behaviors are clearly observed for all positions, and they are assigned as the blinking of q-dots [14,15]. For positions 3, 5 and

7, the blinking shows a single-step behavior, and for position 4 a two-step behavior is observed. The single-step blinking means that the number of q-dot is just one at the position, and the two-step blinking corresponds to the two q-dots at the position. For other positions 1, 2, and 8, the single-step blinking behaviors were observed, but for the position 6 two-step blinking was observed. Regarding the position 8, the off-period was quite long compared to the on-period, and we could not obtain reliable data from the position 8. In the followings, we describe the results for the positions from 1 to 7.

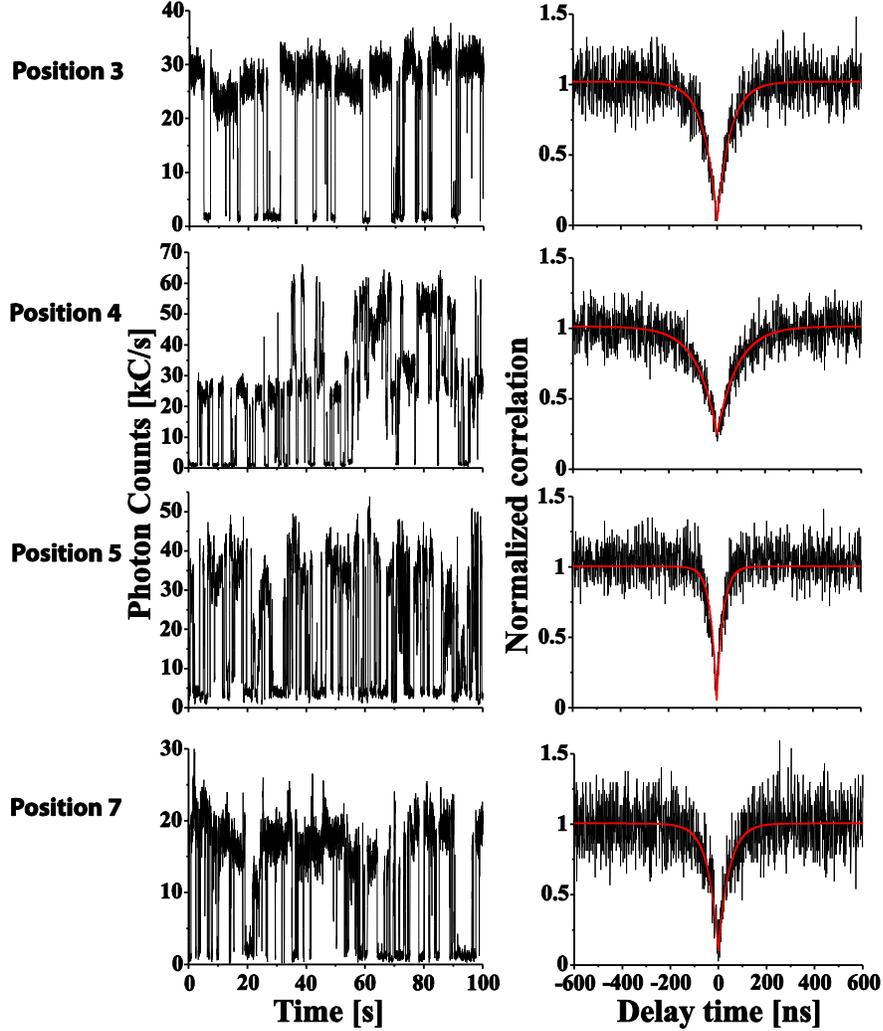

Fig. 3. (Color Online) The left column shows the photon counts as a function of time and the right column shows the normalized photon correlations $g_N^{(2)}(\tau)$ for four positions (3, 4, 5, and 7). Excitation intensity is 50 W/cm$^2$ for the positions of 3, 4, and 5, and 130 W/cm$^2$ for the position 7. The red curves show the exponential fitting of the normalized photon correlations.

The normalized correlations in the right column clearly show anti-bunching behaviors for all positions. The correlations at the anti-bunching dip are close to zero; they are 0.035, 0.034, and 0.05 for positions 3, 5, and 7, respectively, but for position 4, the value is 0.19. Since the normalized correlation for $N$-emitters [16] at $\tau = 0$ is expressed as $g_N^{(2)}(0)=(N-1)/N$, number of q-dots for each position can be estimated to be one for positions 3, 5, and 7. Regarding position 4, we estimate the number to be two, although the dip value is much smaller than the theoretical value of 0.5. This may imply that one dot may be less emissive than the other.

These results are consistent with the blinking behaviors. The dip values are summarized in Table 1. The anti-bunching signals are fitted by single exponential curves (red curves). The obtained recovery times are 55, 83, 25, and 43 ns for the positions 3, 4, 5, and 7 positions, respectively.

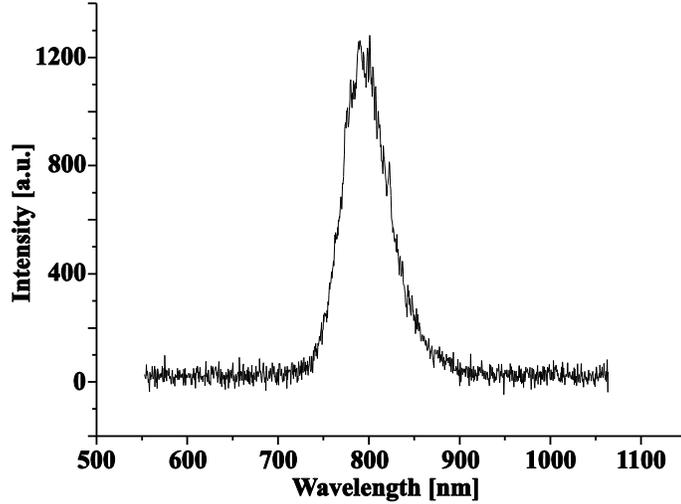

Fig. 4. The emission spectrum for a single q-dot measured at position 3. The excitation intensity is 50 W/cm$^2$. The spectrum reveals, center wavelength is at 796 nm and the spectral width is 52 nm FWHM.

Emission spectrum was measured for all positions. In Fig. 4, typical spectrum for one dot obtained at position 3 is displayed. The spectrum was obtained with an integration time of 5 min with an excitation intensity of 50 W/cm$^2$. The center wavelength ($\lambda_c$) is 796 nm and the spectral width ($\Delta\lambda$) is 52 nm FWHM. Measured $\lambda_c$ and $\Delta\lambda$ values for all the positions are listed in Table 1. Center wavelengths distribute over the range of 80 nm from 746 to 826 nm. Although the measured width is narrower than an ensemble average value of 82 nm [17], it is still broad. The mechanism of the broadening may be understood due to the spectral diffusion and the exciton-phonon interactions which have been discussed for similar q-dots [18,19].

## 4. Analysis

### 4.1 Theoretical model

Figure 5 illustrates the schematic energy-level diagram for fluorescence photon emission process of q-dots. The scheme is essentially an incoherent two-level system, since photo-excited electrons quickly relax to the emission state |1> [20].

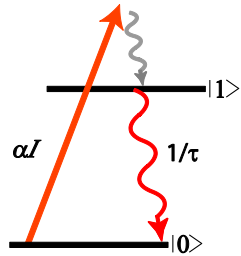

Fig. 5. (Color online) Schematic energy-level diagram for the photo-emission process of q-dots. The gray non-radiative relaxation process is assumed to be very fast.

Excited-state population $N_1$ can be described by the following rate equation,

$$\dot{N}_1 = \alpha I N_0 - \frac{N_1}{\tau},$$

where $N_0$ is the ground-state population, $\alpha I$ is the excitation rate at laser intensity $I$, and $1/\tau$ is the total decay rate of the excited state $|1\rangle$. Assuming single q-dot ($N_0 + N_1 = 1$) and initial condition of $N_1(t=0) = 0$, temporal evolution of the excited-state population is readily derived as follows,

$$N_1(t) = \frac{\alpha I}{\alpha I + 1/\tau}\left\{1 - \exp\left(-\frac{t}{T}\right)\right\} \quad (1)$$

$$1/T = \alpha I + 1/\tau$$

where $1/T$ is the intensity-dependent decay-rate. Then, the photon emission rate into the nanofiber guided modes can be expressed as following.

$$n_{fiber}(t) = N_1(t) \times \frac{1}{\tau_r} \times \eta_c = \frac{\eta_q \eta_c}{\tau}\frac{\alpha I}{\alpha I + 1/\tau}\left\{1 - \exp\left(-\frac{t}{T}\right)\right\} \quad (2)$$

where $1/\tau_r$ denotes the radiative decay rate of the excited state $|1\rangle$ and $\eta_c$ the coupling efficiency of the fluorescence photons into the guided modes. It should be mentioned that the radiative decay rate is related to the quantum efficiency ($\eta_q$) of q-dot by $\eta_q = (1/\tau_r)/(1/\tau)$.

The observable photon-count rate by the APD ($n_{obs}$) can be obtained from the photon-emission rate ($n_S$) under the stationary condition by multiplying the transmission efficiency $\kappa$ determined in Section 2 and the quantum efficiency ($\eta_{APD}$) of APD. $n_S$ and $n_{obs}$ can readily be obtained from Eq. (2) as following. A factor 1/2 is introduced due to the fact that the coupled photons are detected only for one direction of the nanofiber. Note that the rates are dependent on the excitation laser intensity.

$$n_{obs}(I) = \eta_{APD}\kappa\frac{n_S(I)}{2} = \eta_{APD}\kappa\frac{\eta_q \eta_c}{2\tau}\frac{\alpha I}{\alpha I + 1/\tau} = n_{obs}(\infty)\frac{\alpha I}{\alpha I + 1/\tau} \quad (3)$$

*4.2 Data analysis*

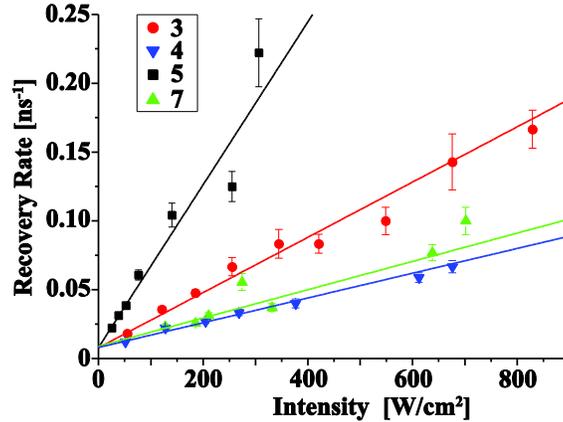

Fig. 6. (Color online) The observed anti-bunching recovery rates ($1/T$) for different excitation laser intensities at positions 3 (red circles), 4 (blue triangles), 5 (black squares) and 7 (green triangles). The solid lines show the linear fits to the data.

Since the anti-bunching recovery process is essentially a population evolution of the excited state, the recovery rates can be expressed by the intensity-dependent decay-rate $1/T$ in Eq. (1). In Fig. 6, observed recovery rates for positions 3, 4, 5, and 7 are plotted versus excitation laser intensity. One can readily see the linear dependence. The data points are fitted by the least-mean square method, and the fitted results are shown by solid lines. As seen in Eq. (1), the intercept at the zero-intensity gives the decay rate of the excited state $1/\tau$. Obtained decay times are listed in Table 1. The decay times distribute in the range of 150±50 ns for the seven positions.

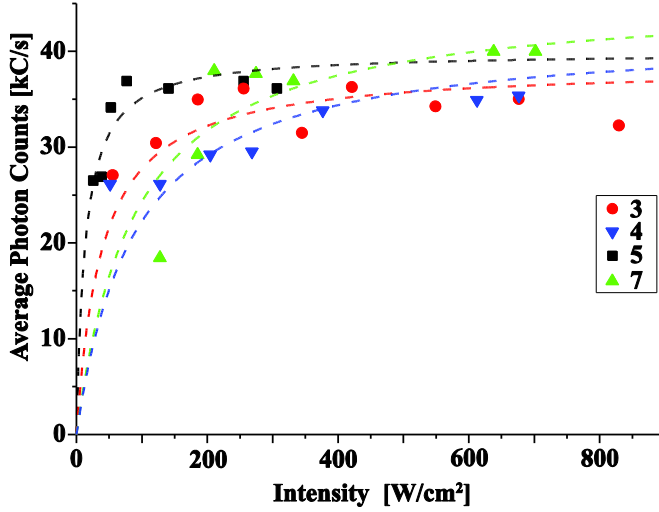

Fig. 7. (Color online) The observed fluorescence photon-count rate for different excitation intensities at the four positions 3 (red circles), 4 (blue triangles), 5 (black squares) and 7 (green triangles). The observed photon counts are fitted (dashed curves) using Eq. (3).

Figure 7 shows the fluorescence photon-count rate versus excitation intensity measured for the four positions 3, 4, 5, and 7. One can readily see the saturation behaviors for all plots. We fitted the measured results with Eq. (3). We fixed the parameters $\alpha$ and $1/\tau$ to the values determined from the intensity dependence of the anti-bunching recovery rates. The adjustable parameter is $n_{obs}(\infty)$. The fitted results are shown by dashed curves and the obtained parameter $n_{obs}(\infty)$ is listed in Table 1 for the seven positions. We estimate $\eta_q \eta_c$ from the obtained parameters using the relation $n_{obs}(\infty) = \eta_{APD} \kappa \eta_q \eta_c / 2\tau$, where $\eta_{APD}$ is assumed to be 60%. The obtained values are shown in Table 1.

**Table 1. Obtained parameters for q-dots on an optical nanofiber.**

| Position | 1 | 2 | 3 | 4 | 5 | 6 | 7 |
|---|---|---|---|---|---|---|---|
| $g_N^{(2)}(0)$ | 0.042 (±0.002) | 0.031 (±0.002) | 0.035 (±0.002) | 0.190 (±0.001) | 0.034 (±0.001) | 0.450 (±0.014) | 0.050 (±0.002) |
| $\lambda_c$ [nm] | 798 | 820 | 796 | 784 | 798 | 826 | 746 |
| $\Delta\lambda$ [nm] | 58 | 68 | 52 | 52 | 60 | 72 | 52 |
| $\tau$ [ns] | 100 (±40) | 195 (±90) | 130 (±20) | 126 (±13) | 140 (±45) | 190 (±50) | 130 (±50) |
| $n_{obs}(\infty)$ [kC/s] | 30.0 (±1.0) | 20.0 (±0.8) | 38.4 (±1.2) | 42.0 (±2.5) | 39.8 (±0.9) | 44.7 (±5.6) | 45.7 (±2.2) |
| $\eta_q \eta_c$ | 0.033 (±0.014) | 0.043 (±0.020) | 0.054 (±0.010) | 0.057 (±0.009) | 0.060 (±0.021) | 0.094 (±0.030) | 0.064 (±0.028) |

## 5. Discussion

Table 1 summarizes the measured results for the seven positions. The second row shows the $g_N^{(2)}(0)$-values. One can readily see that single q-dot is deposited for positions 1, 2, 3, 5, and 7 where $g_N^{(2)}(0) \ll 1$. On the other hand, two q-dots are deposited for positions 4 and 6. For position 4, the situation is described in Section 3. Regarding position 6, $g_N^{(2)}(0)$-value is close to 0.5. This means that the deposited two q-dots behave almost the similar to each other. This situation may show a good contrast to that for position 4. The observations reveal that single q-dot can be deposited reproducibly using the present method. It would be meaningful to mention the success probability to deposit single q-dot on nanofiber. In average, the probability is estimated to be about 60%.

As seen for the third row to sixth row of Table 1, the obtained parameter values show some variations for the seven positions. These variations would be reasonable, since in the present experiments, we picked up only one q-dot or two q-dots at each position from a huge amount of q-dots in the q-dot solution, and the physical properties of individual q-dot should show some variations around a mean value.

It would be meaningful to estimate the coupling efficiency of emitted photons into the nanofiber from the derived $\eta_q\eta_c$-values shown in the seventh row of Table 1. The $\eta_q\eta_c$-values show a variation from 0.033 to 0.094. We suspect that this variation may directly reflect the variation of the quantum efficiency of q-dots, since the $\eta_c$-value can be assumed to be kept equal for the seven positions which distribute along the nanofiber within the range of 150 μm. Regarding the quantum efficiency $\eta_q$, the company quotation is 72% [17,21,22]. It should be noted that the value is a mean value measured for a liquid sample which contains a huge amount of q-dots. The $\eta_q$-value for single q-dot may distribute around the mean value. Hence, the obtained highest $\eta_q\eta_c$-value may correspond to a q-dot which has the highest quantum efficiency among the measured q-dots. Assuming the highest $\eta_q$-value to be 100%, we estimate the $\eta_c$-value. Note that the value gives the lower limit for the coupling efficiency, since the $\eta_q$-value of 100% should be an overestimated value. Thus, for the obtained highest $\eta_q\eta_c$-value of 0.094 at the position 6, we have estimated the lower limit of $\eta_c$-value to be 9.4±3%.

The coupling efficiency has been theoretically calculated as 22% assuming the randomly oriented dipoles [4,5]. If we use the theoretical $\eta_c$-value, the resulting $\eta_q$-values are all much smaller than the mean value of 72%. It may suggest that the experimental results imply a smaller coupling efficiency than the theoretically calculated value. To clarify this point, further investigations such as absolute measurement of the $\eta_c$-value would be necessary.

## 6. Summary

We have experimentally investigated the fluorescence photon emission characteristics for single q-dots by using an optical nanofiber. We have demonstrated that single q-dots can be deposited along an optical nanofiber systematically and reproducibly within a precision of 5 μm. Such a technique can be promising for depositing single colloidal q-dots in various micro-resonators and optical nanofiber cavity. For single q-dots on an optical nanofiber, we measure the fluorescence photon numbers coupled into the nanofiber and the normalized photon correlations, by varying the excitation laser intensity. We have estimated the fluorescence photon coupling efficiency into the nanofiber guided modes to be higher than 9.4±3%. This work would give a basis for extending the optical nanofiber method to a general technique in quantum optics and quantum information sciences.

### Acknowledgements

R. R. Yalla acknowledges the support of a Japanese Government (Monbukagakusho) scholarship. This work was carried out as a part of the Strategic Innovation Project by Japan Science and Technology Agency.